**What future for the Anthropocene? A biophysical interpretation.**
By Ugo Bardi

Dipartimento di Scienze della Terra, Università di Firenze, Italy.
Polo Scientifico di Sesto Fiorentino, 50019 Sesto F. (Fi) Italy
ugo.bardi@unifi.it

*Abstract. The Anthropocene is a proposed time subdivision of the earth's history correlated to the strong human perturbation of the ecosystem. Much debate is ongoing about what date should be considered as the start of the Anthropocene, but much less on how it can evolve in the future and what are its ultimate limits. It is argued here that the phenomena currently defining the Anthropocene will rapidly decline and disappear in times of the order of one century as a result of the irreversible dispersal of the thermodynamic potentials associated to fossil carbon. However, it is possible that, in the future, the human economic system may catalyze the dissipation of solar energy in forms other than photosynthesis, e.g. using solid state photovoltaic devices. In this case, a strong human influence on the ecosystem may persists for much longer times, but in forms very different than the present ones.*

The history of the earth system is described in terms of a series of time subdivisions defined by discrete (or "punctuated") stratigraphic changes, mainly in terms of biotic composition (Aunger 2007a) (Aunger 2007b). The most recent of these subdivisions is the proposed "Anthropocene," a term related to the strong perturbation of the ecosystem created by human activity. The starting date of the Anthropocene is not yet officially established, but it is normally identified with the start of the large scale combustion of fossil carbon compounds stored in the earth's crust ("fossil fuels"), and, in this case, it could be located at some moment during the 18$^{th}$ century CE (Crutzen 2002), (Lewis and Maslin 2015). Alternative proposals would place it at the start of the development of agriculture (Ruddiman 2013) (Ruddiman et al 2015), or of nuclear fission technologies (Zalasiewicz et al. 2015). However, it is clear that the human influence on the earth system has enormously increased with the use of fossil fuels and it still remains mainly linked to fossil fuels (Raupach and Canadell 2010). So, the question can be asked of what the evolution of the Anthropocene could be as a function of the decreasing availability of fossil carbon compounds. Will the Anthropocene decline and the earth system return to conditions similar to the previous geologic subdivision, the Holocene? Or will the human perturbation continue? This is the subject examined in the present work.

The earth system is a non-equilibrium system whose behavior is determined by the flows of energy it receives. This kind of system tends to act as energy transducer and to dissipate the available energy potentials at the fastest possible rate (Sharma & Annila 2007), (Kaila and Annila 2008); a property that can also be understood in terms of the principle known as "maximum entropy production" (Kleidon 2004),(Kleidon et al 2010), (Martyushev & Seleznev 2006). Non equilibrium systems tend to attain the property called "homeostasis" if the potentials they dissipate remain approximately constant (Kleidon 2004). In the case of the earth system, by far the largest flow is the energy coming from the sun. It is approximately constant (Iqbal 1983), except for very long time scales, since it gradually increases by a factor of about 10% per billion years (Schroeder and Connon Smith 2008). Therefore, the earth's ecosystem would be expected to reach and maintain homeostatic conditions for very long times. However, this doesn't happen because of geological perturbations that generate the punctuated transitions observed in the stratigraphic record.

The flow of geothermal energy to the earth surface is orders of magnitude smaller than that of the solar energy (Davies and Davies 2010)) and it is known to change very slowly over geological time scales (Korenaga 2008). However, it is subjected to punctuated changes in the form of volcanic eruptions and tectonic movements. These perturbations are the main cause of the transitions between geological time subdivisions. For instance, it is known that during the Phanerozoic period, a strong correlation can be observed between large basaltic eruptions (known as "large igneous provinces", LIPs) and mass extinctions, in turn often associated to geological time boundaries (Kidder and Worsley 2010), (Wignall 2001) (Bond and Wignall 2014). Extraterrestrial factors, such as asteroidal impacts, have also been claimed to generate discontinuities during the Phanerozoic, but their effect may have been overestimated (Archibald 2014), (Bond and Wignall 2014).

A further factor that may create discontinuities in the sedimentary record is linked to evolutionary changes of the biosphere, of the kind defined as "revolutions" (Szathmáry & Smith 1995), (Kleidon 2004), where the ecosystem "learns" how to increase the rate of dissipation of the available potentials (Kaila and Annila 2008). It has been estimated that the amount of solar energy processed by terrestrial organisms has increased by a factor of about one thousand over a time span of about three billion years (Lenton & Watson 2011) (p. 49). Changes in this capability may lead to radical changes in the ecosystem record; for instance, the "great oxygenation event" (GOE) that occurred at ca. 2.5 Ga ago may be related to the Archean/Proterozoic boundary (Gargaud et al. 2011).

The transition that generated the Anthropocene is also related to a discontinuity in the energy dissipation rate of the ecosystem. This discontinuity appeared when the ecosystem (more exactly, the "homo sapiens" species) learned how to dissipate the energy potential stored in the form of the carbon compounds called "fossil carbon" or "fossil fuels" (mainly oil, gas, and coal). These compounds are, in effect, stored solar energy that had slowly accumulated as the result of the sedimentation of organic matter mainly over the Phanerozoic era, that is over a time scale of the order of hundreds of millions of years (Raupach and Canadell 2010). The rate of energy dissipation of this fossil potential, at present, can be estimated in terms of the "primary energy," at the input of the human economic system. In 2013, this amount corresponded to ca. 17 TW (Anon 2014b). Of these, about 86% (or ca. 15 TW) were produced by the combustion of fossil carbon compounds. This is a small amount in comparison, the average total flow of solar energy that reaches the earth's surface is estimated as 89,000 TW (Tsao et al. 2006) or 87,000 TW (Szargut 2003). However, the energy directly produced by combustion is just a trigger for other, more important effects. Among these, we may list as the dispersion of large amounts of heavy metals and radioactive isotopes in the ecosphere, the extended paving of large surface areas by inert compounds (Schneider et al 2009), the destruction of a large fraction of the continental shelf surface by the practice known as "bottom trawling" (Zalasiewicz et al 2011), and more. The most important effects are related to the emission of greenhouse gases as combustion products, mainly carbon dioxide, $CO_2$, but also other gases (Stocker et al 2013). The thermal forcing generated by $CO_2$ alone can be calculated as approximately 900 TW, or about 1% of the solar radiative effect (Zhang and Caldeira 2015), hence a non negligible effect that generates an already detectable greenhouse warming of the atmosphere. This warming, together with other effects such as oceanic acidification, has the potential of deeply changing the ecosystem in the same way as, in ancient times, LIPs have generated mass extinctions and major changes of the ecosystem (Wignall 2005) (Bond and Wignall 2014).

The emission of greenhouse gases in the atmosphere will continue as long as the combustion of fossil fuels continues. However, carbon compounds exist in limited amounts inside the earth's crust. The total mass of fossil carbon is estimated as ca. $1.5 \times 10^{16}$ t ($1.25 \times 10^{21}$ mol C), mainly in the form of the family of compounds known as "kerogen" (Falkowski et al 2000). Considering that the present combustion rate of fossil carbon is reported to be about $9.2 \times 10^{+9}$ t per year (Le Quéré et al 2014) one could conclude that, theoretically, it could continue for more than one million years. But this is obviously impossible for various reasons, first of all because the oxygen in the atmosphere would run out much earlier (estimated as ca $1.2 \times 10^{15}$ t, or $3.7 \times 10^{19}$ mol $O_2$) (Canfield 2005)). However, a much more drastic limitation derives from the fact that not all the fossil carbon present in the earth's crust is "burnable" carbon; that is carbon that will likely be burned by the human industrial system.

The quantity of burnable carbon is normally assessed on the basis of factors related to the human economy; that is, in term of the carbon compounds whose cost of extraction can provide a profit at the current prices of energy. A comprehensive estimate based on this concept (Rogner 1997) indicated that the known fossil reserves could theoretically sustain the present consumption rate for about one century and a half, even though "non conventional" resources could extend this period. However, this kind of estimate suffers of fundamental uncertainties since it is impossible to predict what the prices of energy will be in a century from now (and even for much shorter time spans). So, a more reliable estimate should be based on thermodynamic factors.

The combination of atmospheric oxygen with fossil carbon is a chemical reaction and, as such, it needs to overcome a kinetic barrier that we can define as "activation energy." This energy determines the reaction rate;

the higher the barrier, the slower the rate. In the case of the combustion of fossil fuels, the barrier is generated by the fact that fossil carbon is stored underground and it is not, normally, in direct contact with atmospheric oxygen. As a result, during the Phanerozoic period, sedimentation has accumulated carbon faster than it has been removed by natural oxidation or combustion, despite the latter reactions being thermodynamically favored. It is also well known that the activation energy of a chemical reaction can be lowered, and the reaction rate accelerated, by a catalyst. This is the role played today by the human economy: a catalyst that enormously accelerated the oxidation rate of fossil carbon. In addition, the combustion of fossil carbon can be defined as an "autocatalytic" reaction, in the sense that it generates structures that accelerate the reaction rate. This is typical of many chemical reactions and physical processes, which tend to create "dissipative structures," described for the first time by Ilya Prigogine (Prigogine 1967), (Prigogine 1968).

The structures created on the earth by the dissipation of the fossil fuel potentials can be defined with the generic name of "industrial system". Burning fossil fuels generates the exergy needed to create industrial structures which, in turn, are used to extract more fossil fuels and burn them. Described in this sense, the fossil based, human industrial system can be seen as a non-biological metabolic system, akin to biological ones (Malhi 2014). The structures of this non-biological metabolic system can be examined in light of concepts such as "net energy" (Odum 1973) defined as the exergy generated by the transduction of an energy stock into another form of energy stock. Another, similar concept is "energy return for energy invested" (EROI or EROEI), first defined in 1986 (Hall et al 1986) (see also (Hall et al 2014)). EROEI is defined as the ratio of the exergy obtained by means of a certain dissipation structure to the amount of exergy necessary to create and maintain the structure. An EROEI smaller than 1 (or, equivalently, a net energy smaller than zero) implies that the process cannot be self-sustaining. As an example, coal simply burned in air generates no exergy and therefore can't sustain the carbon mining process. Instead, the coal burned in a steam engine generates mechanical energy (exergy) that can be used to manufacture equipment to drill, excavate, lift, and transport more coal; and also to maintain the steam engine and manufacture new ones. If the EROEI associated with a dissipation process is larger than one, the excess can be used to replicate the process in new structures and, on a large scale, to create the complex system that we call the "industrial society." The growth of the human civilization as we know it today, and the whole Anthropocene, can be seen as the effect of the relatively large EROEI associated to the combustion of fossil carbon compounds (Lambert et al 2014).

A peculiarity of the dissipation of potentials associated with fossil hydrocarbons is that the system cannot attain homeostasis. It has been recognized from the times of Stanley Jevons (Jevons 1866) that the mineral industry extracts first the "high grade" resources, defined as the most concentrated ores available. In the case of fossil fuels, the high grade resources are those which produce the maximum exergy for the minimum requirement in exergy for extraction; that is, the maximum EROEI. As the stock of these resources is exhausted, the industry moves to lower quality resources which require more energy expensive extraction structures as, for instance, resources that can only be extracted from large depths or require complex refining processes. This leads to a progressive decline of the EROEI associated with fossil potentials, although in the first phase of the cycle the decline can be reversed by technological factors and scale factors. For instance, Hall and coworkers (Hall et al 2014) show that the EROEI of oil extraction in the United States peaked at around 30 in the 1960s, to decline to values lower than 20 at present. A further factor to be taken into account is called "pollution", which accelerates the degradation of the accumulated capital stock and hence reduces the EROEI of the system as it requires more exergy for its maintenance.

Only a small fraction of the crustal fossil carbon compounds can provide an EROEI > 1, and an even smaller one can provide the large EROEIs that have created the industrial civilization (Hall et al 2009) (Zencey 2013). Therefore, we can expect that the progressive reduction of the average EROEI values will lead to a slowdown of the combustion reaction. Various models describe this phenomenon; historically, the first one goes back to Jevons (Jevons 1866), later on, the concept was taken up from an empirical viewpoint by Hubbert (Hubbert 1956), and later on quantified by several studies, often based on system dynamics (Forrester 1971) (Meadows et al 1972), (Meadows et al 2004), (Bardi and Lavacchi 2009). All these studies show that the cycle of exploitation of a finite energy resource should follow a "bell shaped" curve, although not necessarily a symmetric one. A large number of studies have examined the cycle of exploitation of fossil hydrocarbons with the objective of making quantitative predictions (see, e.g. (Maggio & Cacciola 2012), (Guseo 2011), (Campbell and Laherrere 1998), (Zittel et al 2013). (Turner 2008), (Bardi 2014) ). The results are, obviously, only estimates, but the general conclusion is that the carbon generated power "pulse" of the Anthropocene is destined to last for a few centuries at most, perhaps less than one century.

So, from a geological viewpoint, the active phase of the Anthropocene is destined to last only a very short time. However, the effects on the ecosystem will not disappear: a fraction of the carbon dioxide emitted during the active phase of the Anthropocene may persist in the earth's atmosphere for several tens of thousands of years, even hundreds of thousands (Archer and Caldeira 2009). It is not known how long the carbon dioxide will remain in other crustal reservoirs, such as in the oceans, but in order to see the combustion process completely reversed we should wait for new coal, gas, and oil reservoirs to be reformed; something that may require hundreds of thousands of years, at least. Such reservoirs might never be reformed, in particular those of coal, which are the result of peculiar climate and biological conditions which existed hundreds of millions of years ago, and may never appear again. So, after the active phase of fossil carbon burning, the earth system would not revert to the conditions of earlier ages. The future characteristics of the earth's ecosystem are obviously difficult to determine, but it appears that, apart from truly catastrophic and irreversible changes (Hansen 2007), the earth's climate might remain in an "interglacial" state for at least 10 thousand years in the future (Berger et al) before eventually returning to the cycles of ice ages and interglacials that characterized the Pleistocene. Over a very long run, the gradual increase of the intensity of the solar irradiation will eventually lead to the disappearance of the vertebrates in some 800 million years in the future and, in an even more remote future, to the complete sterilization of the planet one billion and a half years in the future (Franck et al 2006), (Schroeder and Connon Smith 2008).

We may speculate on whether humans could continue to maintain their strong influence on the ecosystem by switching to the dissipation of potentials other than those provided by fossil hydrocarbons. Potentials not related to the sun exist at the earth's surface in the form of geothermal energy (Davies and Davies 2010). and tidal energy (Munk and Wunsch 1998); both are, however, several orders of magnitude smaller than the energy potential generated by solar light. The fission of heavy nuclei (uranium and thorium) is also a non-carbon based energy potential which can be processed and dissipated. However, this potential is limited in extent and cannot be reformed by Earth-based processes. Barring radical new developments, depletion is expected to prevent this process to play an important role in the future (Zittel et al 2013). Nuclear fusion might be a game changer in this respect, but, so far, there is no evidence that the potential associated with the fusion of deuterium nuclei can generate an EROEI sufficient to maintain an industrial civilization, or even to maintain itself.

A different route to maintain the large energy dissipation rate of the Anthropocene could be to process solar energy in a different way than by means of the photosynthetic engine that powers the biosphere. In principle, this is possible. As mentioned before, the flow of solar energy that reaches the earth's surface is estimated as 89,000 TW (Tsao et al 2006) or 87,000 TW (Szargut 2003). The atmospheric circulation generates some 1000 TW of kinetic energy (Tsao et al 2006). These are orders of magnitude larger than the flow of primary energy associated to the Anthropocene (ca. 17 TW). Of course, as discussed earlier on, the capability of a transduction system to create complex structures depends on the EROEI of the process. On this point, all the recent studies on photovoltaic systems report EROEIs larger than one for the production of electric power by means of photovoltaic devices (Rydh and Sandén 2005), (Richards and Watt 2007), (Weißbach et al 2013). (Blankenship et al 2011), (Chu 2011), (Bekkelund 2013) (Prieto and Hall 2011). In most case the EROEI of PV systems is reported to be smaller than that of fossil burning systems, but, in some cases, it is reported to be larger (Raugei et al. 2012), with even larger values being reported for CSP (Montgomery 2009), (Chu 2011). Overall, values of the EROEI of the order of 5-10 for direct transduction of solar energy can be considered as reasonable estimates (Green and Emery 2010). Even larger values of the EROEI are reported for wind energy plants (Kubiszewski et al 2010). These values may increase as the result of technological developments, but also decline facing the progressive occupation of the best sites for the plants and to the increasing energy costs related to the depletion of the minerals needed to build the plants.

The long term sustainability of solar and wind technologies is a complex subject that can't be discussed in detail here; it suffices to say that the current photovoltaic and wind technologies may use, but do not necessarily need, rare elements that could face near term exhaustion problems (García-Olivares et al 2012). Studies have also reported that the materials used for solar cells can be recycled at rates of 99.99% (Fthenakis 2009). The same result may hold for other technologies, even though the widespread use of rare earths in wind systems makes long term sustainability problematic, at least for the current technology. In any case, taking into account the future improvements for these technologies, there do not appear to exist fundamental barriers to "close the cycle" and to use the exergy generated by human made solar powered devices (in particular PV systems) to recycle the systems for a very long time.

Various estimates exist on the ultimate limits of energy generation from photovoltaic systems. The "technical potential" in terms of solar energy production of the US alone is estimated as more than 150 TW (Lopez et al 2012). According to the data reported in (Liu et al. 2009), about 1/5 of the area of the Sahara desert (2 million square km) could generate around 50 TW at an overall PV panel area conversion efficiency of 10%. Summing up similar fractions of the areas of major deserts, PV plants (or CSP ones) could generate around 500-1000 TW, possibly more than that, without significantly impacting on agricultural land. In addition, wind energy could generate as much as about 80 TW, (Jacobson & Archer 2012), or somewhat smaller (Miller et al. 2011) values, although perhaps no more than 1 TW (de Castro et al. 2011). Overall, these values are much larger than those associated with the combustion of fossil fuels, with the added advantage that renewables such as PV and wind produce higher quality energy in the form of electric power.

From these data, we can conclude that the transduction of the solar energy flow by means of inorganic devices could represent a new metabolic "revolution" of the kind described by (Szathmáry and Smith 1995). (Lenton and Watson 2011) that would bootstrap the ecosphere to a new and higher level of transduction. It is too early to say if such a transition is possible, but, if it were to take place at its maximum potential, its effects could lead to transformations larger than those associated to the Anthropocene as it is currently understood. These effects are hard to predict at present, but they may involve changes in the planetary albedo, in the weather patterns, and in the general management of the land surface. It is clear that when dealing with such levels of energy generation, the transition should be carefully managed to maintain a healthy planetary ecosystem. Overall, the effect might be considered as a new geological transition. As these effects would be mainly associated to solid state devices (PV cells), perhaps we need a different term than "Anthropocene" to describe this new phase of the earth's history. The term "*stereocene*" (the age of solid state devices) could be suitable.